\documentclass[amsmath,prl,aps,twocolumn]{revtex4}
\usepackage{amsthm,amsfonts,verbatim} 
\usepackage{hyperref,epsfig}

\def\Tr{\qopname \relax o{Tr}}
\def\Vol{\qopname \relax o{Vol}}
\newcommand{\tr}{{\rm tr}}

\newcommand{\D}{{\rm d}}

\def\RR{{\mathbb R}}
\def\SS{{\mathbb S}}

\begin{document}

\title{Entanglement entropy of fermions in any dimension and the Widom
conjecture}
\author{Dimitri Gioev$^{1}$}%
\email{gioev@math.rochester.edu}
\author{Israel Klich $^{2}$ }%
\email{klich@caltech.edu}
\affiliation{(1) Courant Institute, New York University, New York, NY 10012 \\
and Department of Mathematics, University of Rochester,
Rochester, NY 14627 \\
(2) Department of Physics, Caltech, Pasadena, CA 91125}
\begin{abstract}
We show that entanglement entropy of free fermions
scales faster then area law, as opposed
to the scaling $L^{d-1}$ for the harmonic lattice, for example.
We also suggest and provide
evidence in support of an explicit formula for the entanglement entropy of free fermions in any
dimension $d$,
 $S\sim c(\partial\Gamma,\partial\Omega)\cdot L^{d-1}\log L$
as the size of a subsystem $L\to\infty$, where $\partial\Gamma$ is the Fermi surface
and $\partial\Omega$ is the boundary of the region in real space. The expression for
the constant $c(\partial\Gamma,\partial\Omega)$ is based on a conjecture due to
H.~Widom. We prove that a similar expression holds for the particle number fluctuations
and use it to prove a two sided estimates on the entropy $S$.

\end{abstract}
\maketitle

In recent years a number of parallel findings have emphasized the importance of
entanglement entropy [1--10]. Although originally studied in the context of black hole
physics \cite{Bombelli}, this quantity also plays an important role in quantum
information as a measure of the number of maximally entangled pairs that can be
extracted from a given quantum state \cite{Bennet}.

The behavior of entanglement entropy is closely related to the criticality behavior of
quantum systems: for a gapped system one expects an area scaling law due to a finite
correlation length $\xi$.
In the 1D case \cite{Vidal}, this behavior changes drastically near criticality where
the absence of gap leads to long range correlations, and so the entanglement does not
saturate. Many interesting results have been obtained for
1D models. 
 For classes of critical models, where
conformal field theory (CFT) methods are applicable, the entropy was found to exhibit a
logarithmic behavior, with a coefficient depending on the central charge of the CFT
models \cite{Cardy,CallanWilczek}, recently a modification of these results in case of
strong disorder was found \cite{Refael Moore}.
The bi-partite structure of the ground
state of Fermion models was studied in several works \cite{Peschel Botero
Levay,klich,wolf}. A connection between the entropy of spin chains and Random Matrix
Theory was established in \cite{Keating}.

Fewer results were obtained in dimensions $d>1$, although from the field theoretic
point of view these are very interesting. Indeed, initial investigation of the entropy
as a correction to the Bekenstein-Hawking entropy, suggested that the entropy of a
scalar field is proportional to the boundary area for spherical or a half-space regions
\cite{Bombelli,Srednicki}. Recently, it has been rigorously proved \cite{Plenio} for a
harmonic lattice model, that the entropy of a cube with side $L$
 behaves as the boundary area, i.e.~as $L^{d-1}$.

In this Letter we examine the dependence of entanglement entropy on dimension and
geometry in a simple case of a gapless system consisting of free fermions.

Let us summarize our main results: First, we prove that $S\sim L^{d-1}\log
L$ \footnote{$\log$ denotes $\log_2$ and $\ln$ denotes $\log_e$.} for cube-like domains
(\eqref{twosided} below). We then present a heuristic argument for the more explicit
formula $S\sim {\frac{d}{3}}L^{d-1}\log{L}$. We note that results which are derived for
cubes do not necessarily describe the scaling for general boundaries: indeed we find
that $S\geq L^{d-\beta}$ for fractal-like boundaries, where $\beta\in(0,1)$ (described
below) characterizes regularity of the boundary. However, the results for cube-like
domains should reflect the correct scaling for regions with sufficiently regular
boundaries.
For general piecewise smooth
boundaries we prove $O(L^{d-1}\log L)\leq S \leq
O(L^{d-1}(\log L)^2)$, see \eqref{entropy up and low bound},
\eqref{thmspec} (this estimate was independently
derived in \cite{wolf} for $d$ dimensional
cubes in the lattice case).

Finally, making a connection with a conjecture of Widom \cite{Widom82}, we suggest an explicit
geometric formula for the entropy as $L\rightarrow\infty$:
\begin{eqnarray}\label{entr_conj}
S \sim \frac{{L}^{d-1}\log{L}}{(2\pi)^{d-1}} \,\frac1{12}
\int_{\partial\Omega}\int_{\partial\Gamma} |n_x\cdot n_{p}| \D S_x\D S_{p},
\end{eqnarray}
where $\partial\Gamma,\partial\Omega$ are the boundaries of the Fermi sea and the
region considered, $n_p,n_x$ are the unit normals to these boundaries.
We present
evidence supporting this conjecture and prove a similar formula for the fluctuations in
particle number in the subsystem, which also gives bounds on $S$.
Recently the formula \eqref{entr_conj} was checked numerically
for 2D and 3D \cite{num_check}
and a perfect agreement concerning both the order and the coefficient
was found.
Widom's conjecture is
closely related to the problem of recovering data from a measurement during a finite
time interval and in a finite frequency set. This problem, known as {\em time-frequency
limiting,\ }is of basic importance in signal theory, and it was studied extensively
\cite{timefreqlimiting}.
It turns out that operators appearing in calculations of entanglement entropy for free
fermions are exactly the same as the ones studied in \cite{timefreqlimiting}, which is
natural since one studies the properties of a field in restricted sets of real space
and momentum space.

\begin{figure}[tb]
\begin{center}
\epsfig{file=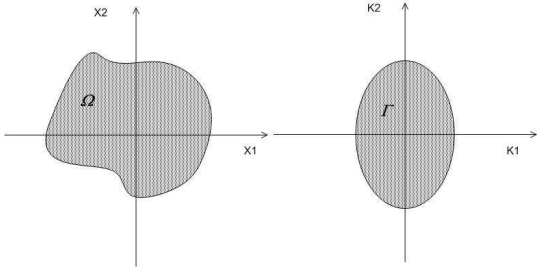,scale=0.35}
\caption{The Fermi sea $\Gamma$ in momentum
space, and a region $\Omega$ in real space}
\label{regions}
\end{center}
\end{figure}

The ground state of a translation invariant Hamiltonian describing a non-interacting
fermion field (on a lattice or in the continuum),
with dispersion relation $\epsilon(k)$, is
$\prod_{\epsilon(k)\leq\epsilon_F}a^{\dag}_k|0>$. Here $\epsilon_F$ is the Fermi energy.
This defines the Fermi sea region $\Gamma=\{k|\epsilon(k)\leq\epsilon_F\}$ in momentum
space. We also assume that the system is gapless \footnote{In the lattice case the
Fermi energy is assumed within the conduction band.}. The bi-partite structure of the
ground state can be studied by fixing a region $\Omega$ in real space and computing the
reduced density matrix $\rho_{\Omega}=\Tr_{({\cal{F}}( \mathbb{R}^d \backslash
{\Omega}))}\rho$ where ${\cal{F}}({\Omega})$ is the fermion Fock space associated with
the region ${\Omega}$ (see Fig.~\ref{regions}). The entanglement entropy
$S=-\Tr_{({\cal{F}}({\Omega}))}\rho_{\Omega}\log \rho_{\Omega}$ is given in this case
\cite{klich} by
$   S({L})=\Tr h(PQP)$
where $h=h_1+h_2$ with $h_1(t)=-t\log t$ and $h_2(t)=-(1-t)\log(1-t)$. Here $P$ is a projection
operator on the modes inside the Fermi sea $\Gamma$, and $Q$ is a projection on the region
${\Omega}$ scaled by a factor ${L}$. The operator $PQP$ is related to the fermion correlation
function $g(x-x')=<a^{\dag}_x a_{x'}>=<x|P|x'>$ so that $<x|PQP|x'>=\int_{\Omega}
g(x-x'')g(x''-x')\D x''$. The density of particles is $n=vol(\Gamma)$, and one may rescale ${L}$
appropriately, as to set $n=1$, which we will assume from now on.

{\it Results for cubic domains.--} Consider the case of a rectangular box with sides
${L}_j$ i.e. $\Omega=[0,L_1]\times\cdots\times[0,L_d]$ and $\Gamma=[0,1]^d$. Let
$S_1({L})$ be the entropy in the 1D case and let $S$ be the entropy corresponding to
$\Omega,\Gamma$ as above. Then we have the following

{\it Theorem.} Under the above assumptions
\begin{eqnarray}\label{thm}
\begin{aligned}
{\frac1{2d}}\sum_{j=1}^d S_1({L}_j)\prod_{i\neq j}N({L}_i)&\leq S
  \leq \sum_{j=1}^d S_1({L}_j)\prod_{i\neq
j}N({L}_i)
\end{aligned}
\end{eqnarray}
where $N(L_j)$ is the average number of particles. Note, in particular, that for
$\Omega=[0,L]^d$
\begin{eqnarray}\label{twosided}
{\frac12}\Big({L\over 2\pi}\Big)^{d-1} S_1(L)\leq S \leq d \Big({L\over
2\pi}\Big)^{d-1} S_1(L).
\end{eqnarray}

Proof: Note that we can make separation of variables $Q=\otimes_1^{d} Q_j$ where $Q_j$
is a projection on coordinate $j$, and $P$ factors in a similar way. Hence $PQ
P=\otimes_1^{d}T_j$ where $T_j=P_jQ_{j}P_j$. Note the following

{\it Lemma.} For $a_i\in[0,1]$ one has
\begin{eqnarray}\label{ineq}&
{\frac1{2d}}G(a_1,\cdots,a_d)\leq h_2(\prod_{i=1}^d a_i)\leq
G(a_1,\cdots,a_d)
\end{eqnarray}
where $G(a_1,\cdots,a_d)=\sum_{j=1}^d h_2(a_j)\prod_{i\neq j}a_i$.
To prove \eqref{ineq}, one has to check it for two variables
$a_1,a_2$ and then proceed by induction.

We observe that the eigenvalues of $PQP$ are of the form $
a_{i,1}\cdots a_{i,d}$ with $a_{i,j}$ being some eigenvalue of
$T_j$. Writing the entropy $S$ 
as $\sum
h(a_{i,1}\cdots a_{i,d})$, using \eqref{ineq} and
$h_1(\prod_{i=1}^d a_i)= \sum_{j=1}^d h_1(a_j)\prod_{i\neq j}
    a_i
$, and recalling that the average number of particles \cite{klich}
$N=\Tr T_j$, \eqref{thm} follows.

Result \eqref{thm} shows that the entropy may be evaluated using the 1D expressions.
This reflects the compatibility of $\Gamma,\Omega$ with factorizing
the fermionic modes into the different coordinates. It is now a matter of substituting
the numerous results obtained in the 1D case. For the lattice case, it follows from the
many works on the subject that for fermions on a 1D lattice, or equivalently for an XX
spin chain (via the Jordan-Wigner transformation), $S_1(L)=\frac13\log L+o(\log L)$, see in particular \cite{JinKorepin}.
 In the continuous case the same expression is obtained by formally
substituting $h(t)$ in the 1D result of \cite{LandauWidom80} \footnote{We believe the last
argument can be made rigorous, but that is outside the scope of the present
paper. The function
$\frac{h(t)}{t(1-t)}$ is not {\it Riemann }integrable
and hence the result of \cite{LandauWidom80} is not applicable as such.}.

For any body composed of a union of cubes $C_i$ of side ${L}_i$ we have, using the subadditivity of
entropy \cite{Lieb Ruskai}, $S(\bigcup C_i)\leq \sum S(C_i)$, thus we have an upper bound that
depends on the number of cubes needed to describe the body: using \eqref{twosided} we find
 $S(\bigcup
C_i)\leq {\frac{d}3 (\frac1{2\pi})^{d-1}}\sum{L}_i^{d-1}\log({L}_i)+o(L_M^{d-1}\log L_M)$, $L_M=\max_i
L_i$. A lower bound proportional to $L_m^{d-1}\log L_m$, $L_m=\min_i L_i$
 follows from \eqref{entropy up and low bound}, \eqref{thmspec} below.

{\it Scaling coefficient.--} 
Here we derive heuristically
\begin{eqnarray}\label{entr_cube}
  S={\frac{d}3}\Big(\frac{L}{2\pi}\Big)^{d-1}\log{L}
     + o({L}^{d-1}\log{L})
\end{eqnarray}
for $\Omega=\Gamma=[0,L]^d$.
Since the eigenvalues of $T_j$ are strictly
less than one, the series
$\Tr h_2(PQP)=\Tr(\otimes_1^d T_j)-\sum_{n=2}^\infty
 {\frac1{n(n-1)}}\Tr(\otimes_1^d T_j^n)
$ 
converges.
By
\cite{LandauWidom80} (see \eqref{Wconj} below for $d=1$ and
$f(t)=t^n$), as $L\to\infty$,
$
   \Tr T_j^n=  \frac{L}{2\pi} +\frac{\log{L}}{\pi^2} \sum_1^n
{\frac1k}+o(\log{L}) 
$. 
Hence $\Tr(\otimes_1^d T_j^n)=
(\frac{L}{2\pi})^d+(\frac{L}{2\pi})^{d-1}\frac{\log{{L}}}{\pi^2}
 (\sum_1^n {\frac1k})
+o(L^{d-1}\log L)$. Substituting the latter in the series for $h_2$
and calculating the sums involved we find:
$\Tr h_2(PQP)={\frac{d}6} (\frac{L}{2\pi})^{d-1}\log{L} + o({L}^{d-1}\log{L})$. Adding this to $\Tr
h_1(PQP)$ which is computed directly and gives the same value, \eqref{entr_cube} follows
\footnote{Note that \eqref{entr_cube} coincides with the upper bound in \eqref{twosided}.}.
Further control of the remainder
term in $\Tr T_j^n$ as $L\to\infty$ 
 is required to make this
calculation rigorous \footnote{In order to justify the interchange of the asymptotic limit in
${L}$, and the series expansion.}.

{\it Results for general boundaries.--} We now turn to the case of general bounded Fermi sea
$\Gamma$ and region $\Omega$.
It is known \cite{klich} that the variance in particle number, given
by $(\Delta N)^2=\Tr PQP(1-PQP)$ can be used to obtain a lower bound on $S$.

{\it Theorem.\ }For general sets $\Omega,\Gamma$ one has
\begin{eqnarray}\label{entropy up and low bound} 4(\Delta N)^2\leq
S\leq O(\log L) (\Delta N)^2.
\end{eqnarray}
 We derive also an explicit
formula \eqref{thmspec} which implies in particular that $(\Delta
N)^2=O(L^{d-1}\log L)$.

The proof of \eqref{entropy up and low bound} in the lattice case is immediate using the
inequalities of the form $4t(1-t)\leq h(t)\leq \epsilon-C t(1-t)\log\epsilon $, valid for
$\epsilon>0$, with $C$ being a constant \cite{Fannes}. One substitutes the operators $PQP$ instead
of $t$ and calculates the trace. Note that $\tr (\epsilon)\sim \epsilon L^d$ for a finite
lattice of size $L$, thus taking $\epsilon<{\frac{\log{L}}{L}}$, \eqref{entropy up and low bound}
follows. The proof of \eqref{entropy up and low bound} in the continuous case is new: note first
that the kernel of the operator $PQP$ is given by
\begin{eqnarray}
<p|PQP|p'>=\chi_\Gamma(p)\chi_\Gamma(p')\Big(\frac{{L}}{2\pi}\Big)^d\int_{\Omega}e^{i{L}(p-p')\cdot
x}\D x
\end{eqnarray}
where $\chi_{A}$ is defined for any set $A$ as $\chi_A(x)=1$ if $x\in A$ and $\chi_A(x)=0$
otherwise. For the continuous case the mentioned inequality \cite{Fannes} is not helpful, since the
Hilbert space associated with any set $\Omega$ is infinite dimensional, so $\Tr \epsilon=\infty$ .
We proceed as follows: write instead
\begin{eqnarray}
h(t)\leq \epsilon\sqrt{t(1-t)}-C t(1-t)\log\epsilon
\end{eqnarray}
valid for small enough $\epsilon$ (with a different constant $C$).
We then take trace of both sides. It remains to estimate $\Tr
\sqrt{PQP(1-PQP)}$. We have the following: $\Tr
\sqrt{PQP(1-PQP)}\leq\Tr \sqrt{PQ_{{\Box}}P}$ where $Q_{{\Box}}$ is a
projection on a box containing $\Omega$, and we have used operator
monotonicity of $t\rightarrow t^{1/2}$ (see e.g.~\cite{pedersen}),
and that $PQ_{{\Box}}P\geq PQP$ (as operators). Next we note that
the operators $PQ_{{\Box}}P$ and $Q_{{\Box}}PQ_{{\Box}}$ have the
same positive eigenvalues counted with multiplicities
\footnote{This holds since for any positive trace class operators
$A,B$ the fact that $\Tr A^n=\Tr B^n$ for all $n=1,2,\cdots$
implies that $A$ and $B$ have the same positive eigenvalues.}.
Thus we have $\Tr \sqrt{PQP(1-PQP)}\leq\Tr
\sqrt{Q_{{\Box}}PQ_{{\Box}}}\leq\Tr
\sqrt{Q_{{\Box}}P_{{\Box}}Q_{{\Box}}}$ where we have used the
monotonicity again. It remains to evaluate $\Tr
\sqrt{Q_{{\Box}}P_{{\Box}}Q_{{\Box}}}$ this can be done using
bounds on the singular values of the operator
$Q_{{\Box}}P_{{\Box}}Q_{{\Box}}$, which in this case are also the
respective eigenvalues. It follows from \cite{BirmanSolomyak} that
for any $\alpha>d/2$ the eigenvalues of
$Q_{{\Box}}P_{{\Box}}Q_{{\Box}}$ satisfy $\lambda_n\leq C
n^{-1/2-{\alpha/d}}L^{d/2+\alpha}$. Taking $\alpha>3d/2$ we
find $\Tr \sqrt{Q_{{\Box}}P_{{\Box}}Q_{{\Box}}}<L^{d+\delta}$ for
any $\delta>0$, and thus we can choose e.g. $\epsilon<L^{-\delta-1}$,
 and \eqref{entropy up and low bound} follows.

Having established $(\Delta N)^2$ as a way of obtaining bounds we
proceed to evaluate it. Our next result is
%

{\it Theorem.}  Let $\Omega,\Gamma$ be two compact sets in
$\RR^d$, $d\geq1$, with smooth boundaries
$\partial\Omega,\partial\Gamma$. Then
\begin{eqnarray}\label{thmspec}
\begin{aligned}&
(\Delta N)^2=\frac{{L}^{d-1}\log{L}}{(2\pi)^{d-1}}
\,\frac{\ln2}{4\pi^2}
     \int_{\partial\Omega}\int_{\partial\Gamma}
          |n_x\cdot n_{p}| \D S_x\D S_{p}\\ &+ o({L}^{d-1}\log{L})
\end{aligned}
\end{eqnarray}
where $n_x,n_{p}$ are unit normals to
$\partial\Omega,\partial\Gamma$, respectively. The full proof is
too technical to be included here, and will appear
 elsewhere (see however \cite{dgioev}). It
starts by observing that
\begin{eqnarray}\label{trtwo}
\Tr (PQP)^2 = \Big(\frac{L}{2\pi}\Big)^{2d} \int_{\RR^d}
A_\Omega(z)\hat{A}_\Gamma({L} z) \,dz
\end{eqnarray}
where
$A_\Omega(z)\equiv\int_{\RR^d}\chi_\Omega(x)\chi_\Omega(x-z)\,dx$
is the volume of the set $\Omega$ intersected with $\Omega$
shifted by $z$ (i.e. $\Omega\cap(\Omega+z)$), and proceeds with an
asymptotic analysis of this integral.

Note the geometric nature of the coefficient in \eqref{thmspec}: for a spherical Fermi sea
$\Gamma$, and a convex region $\Omega$ this coefficient is just the average cross--section of
$\Omega$ over all directions. In the more general case the coefficient depends on the two surfaces
and their mutual orientations. Thus \eqref{entropy up and low bound} with \eqref{thmspec} establish
the scaling $O(L^{d-1}\log L)\leq S \leq O(L^{d-1}(\log L)^2)$
when 
$\Gamma, \Omega$ have smooth boundaries.

{\it Fractal Boundaries.--} A very interesting 
enhancement in the scaling of $S$ occurs if the sets $\Omega,\Gamma$ are allowed to have
fractal-like boundaries. Physically, this means that making the boundaries less regular, makes the
typical momentum states more incompatible with the shape of the region $\Omega$, and hence
contribute to enhanced entropy when integrating the external modes. More precisely, it was shown in
\cite{dgioev} that if $C_1 \|h\|^{\beta_\Omega}<{\rm Vol}( \Omega\setminus(\Omega+h))<C_2
\|h\|^{\beta_\Omega}$ for small $\|h\|$ and some $0<\beta_\Omega,\beta_\Gamma\leq1$, and the same
holds for $\Gamma$ with $\beta_\Gamma$, then $(\Delta N)^2$ is bounded above and below by
$\tilde{C}_{1,2}{L}^{d-\min(\beta_\Omega,\beta_\Gamma)}$ if
 $\beta_\Omega\neq\beta_\Gamma$
and $\tilde{C}_{1,2}{L}^{d-\beta_\Omega}\log{L}$ if $\beta_\Omega=\beta_\Gamma$. In particular this
and \eqref{entropy up and low bound} imply that $S>4
\tilde{C}_{1}{L}^{d-\beta_\Omega}\log{L}\,\,{\rm if}\,\, \beta_\Omega=\beta_\Gamma$ and
$S>4\tilde{C}_{1}{L}^{d-\min(\beta_\Omega,\beta_\Gamma)}\,\,{\rm if}\,\,
\beta_\Omega\neq\beta_\Gamma$
\footnote{A fractal set similar to the one appearing in
\cite{dgioev} was independently
constructed for $d=1$ in \cite{Fannes}.}.

{\it Connection to Widom's conjecture.--} It turns out that the result \eqref{thmspec}
is a special case of a well-known conjecture by H. Widom (1982). 
The problem of
time--frequency limiting mentioned in the introduction leads to a study of the spectrum
of the operator $PQP$ where $Q$ is a time window scaled by $L$, and $P$ represents a
frequency window. One way of studying the eigenvalues of $PQP$ is to study the
asymptotic behavior of $\Tr f(PQP)$, as
${L}\to\infty$, for some general class of $f$. 
It is conjectured in \cite{Widom82}
 that for a function $f(t)$, analytic on a disc of
raduis $>1$ with $f(0)=0$, the following holds as ${L}\to\infty$
\begin{eqnarray}
\label{Wconj}
\begin{aligned}
 \Tr &f(PQP) = \Big(\frac{L}{2\pi}\Big)^d\,f(1)\int_\Omega\int_\Gamma
       \D x\D{p}\\  &+ \Big(\frac{L}{2\pi}\Big)^{d-1}
               \frac{\ln2\log{L}}{4\pi^2}\,U(f)
     \int_{\partial\Omega}\int_{\partial\Gamma}
          |n_x\cdot n_{p}| \D S_x\D S_{p}\\ &+ o({L}^{d-1}\log{L})
\end{aligned}
\end{eqnarray}
where $n_x,n_{p}$ are unit normals to $\partial\Omega,\partial\Gamma$, respectively, and $U(f)
=\int_0^1 \frac{f(t)-tf(1)}{t(1-t)}\,\D t$. The formula \eqref{Wconj} and a generalized form of it
were proved for $d=1$ in \cite{LandauWidom80} and \cite{Widom82}. For $d\geq2$ only special cases
were proved \cite{Widom90,dgioev}. Note finally that \eqref{thmspec} is a verification of Widom's
conjecture for the special case $f(t)=t(1-t)$.

In a broader context one may think of Widom's conjecture \eqref{Wconj} as a generalization of the
strong (two-term) Szeg\"o limit theorem (SSLT) for the {\em continuous\ }setting. The SSLT plays a
special role in entanglement entropy \cite{JinKorepin}
\footnote{
Indeed, for translation
invariant systems, the correlation matrices are Toeplitz matrices (i.e. $<a^{\dag}_ia_j>=g(i-j)$),
and the asymptotics of Toeplitz determinants are given by (various versions of) the SSLT.}.
The SSLT was initially used by Onsager in his celebrated computation of the spontaneous magnetization for
the 2D Ising model (see e.g.~\cite{Bottcher}). It is interesting to note that in
 Onsager's computation (and also in \cite{JinKorepin}) the leading
asymptotic term vanishes, and one needs to compute the sub-leading term.
This is exactly the
situation that we have in the continuous version of the Szeg\"o theorem \eqref{Wconj}: the leading
term should vanish since $h(1)=0$. 

Widom's conjecture 
suggests the explicit geometric expression for the entropy \eqref{entr_conj}.
%
Note that if $\Omega=\Gamma=[0,1]^d$ then the double integral in \eqref{entr_conj} equals $4d$
(twice the number of faces), so that \eqref{entr_conj} and \eqref{entr_cube} are consistent. Note
also that the coefficient $\frac{\ln2}{4\pi^2}$ in the expression for the number variance
\eqref{thmspec} gives a lower estimate for $S$
in \eqref{entropy up and low bound} 
within $16\%$ of the conjectured $\frac1{12}$ in \eqref{entr_conj}.

{\it Finite temperature.-- }From the semiclassical point of view one expects the
entropy to be extensive, $S\sim L^d$, for $T>0$. This suggests to look for a transition
temperature between the $L^d$ and $L^{d-1}\log L$ regimes. Let $\beta=1/T$ and
introduce the Fermi--Dirac function $k({p})=1/(1+e^{\beta(|{p}|^2-\mu)})$ (we take
$\hbar=k_B=1$ and $m=1/2$). The expression for the entanglement entropy at finite
temperatures \cite{klich} is given by $\Tr h(QKQ)$ where $K$ is the operator of
multiplication by $k({p})$ in momentum space. Semiclassically, integrating over the
phase space one finds \footnote{ In \cite{Widom60} a three-term asymptotics of $\Tr
f(QKQ)$ is proven for analytic $f$. However \eqref{lasteq} strictly speaking does not
follow from \cite{Widom60} since $h$ is not analytic.}
$$
  \Tr h(QKQ) = \Big(\frac{L}{2\pi}\Big)^d \Vol(\Omega)\int_{\RR^d} h(k({p}))\,\D{p}
             + O({L}^{d-1}).
$$
Introducing polar variables and scaling out $\beta$ gives
\begin{eqnarray}\label{lasteq}
\begin{aligned}
   S= &\Big(\frac{L}{2\pi}\Big)^d \Vol(\Omega)|\SS^{d-1}|
            \frac{\mu^{-1+d/2}}\beta \int_{e^{-\beta\mu}}^\infty
             \frac{\D u}{u} \\
    &\times h\Big(\frac1{1+u}\Big)\,
       \Big(1+\frac{\log u}{\beta\mu}\Big)^{-1+d/2}\,
            + O({L}^{d-1})
\end{aligned}
\end{eqnarray}
which scales as ${L}^d\beta^{-1}\mu^{-1+d/2}$ for
$\beta\to\infty$.
Comparing this with the $T=0$ results above we see that for the
zero temperature effect ${L}^{d-1}\log{L}$ to be seen, the
transition temperature should satisfy
$T\mu^{-1+d/2}\sim\frac{\log{L}}{L}$, ${L}\to\infty$.

{\it Summary and Discussion.-- } In systems with finite correlation length $\xi$, one
expects quantities such as the entropy $S$ and the number variance $(\Delta N)^2$ to
scale like the area of the boundary of the region. 
The system studied here does not behave this way. 
Here, the correlation function $<a^{\dag}_xa_{x'}>$ decays slowly and the fermion
momentum modes are spread over the entire system and are highly sensitive to
localization in space and consequently the area law is violated.

Let us summarize the concrete results of this Letter: We prove that the scaling is of
the form $L^{d-1}\log{L}$ for cube like domains. We find a connection between the
scaling behavior of $S$ and a well-known conjecture due to Widom
\eqref{Wconj}, which suggests
the explicit 
geometric formula \eqref{entr_conj} for $S$ in any $d$. Finally,
while Widom's conjecture is far from being proven, we find that
it holds for $(\Delta N)^2$,
and use this to obtain lower and upper bounds on $S$. We also find an enhanced scaling
of $S$ for fractal like boundaries and at finite temperatures.

{\it Acknowledgments.-- }We thank H.~Widom for explaining a result from
\cite{LandauWidom80} and K.~Shtengel
for bringing \cite{wolf} to our attention.
D.~G.~is grateful to Caltech Mathematics Dept.~for
hospitality and support,
and to STINT foundation (Sweden)
for basic support for visiting Caltech.


\end{document}